\documentclass[journal]{IEEEtran}
\usepackage{graphicx}

\usepackage{microtype}
\usepackage{graphicx}
\usepackage{subfigure}
\usepackage{booktabs} 
\usepackage{hyperref}
\usepackage{multicol}
\usepackage{multirow}

\usepackage{wrapfig}

\usepackage{amsmath}
\usepackage{amssymb}
\usepackage{mathtools}
\usepackage{amsthm}
\usepackage{caption}
\usepackage{xcolor}
\usepackage[capitalize,noabbrev]{cleveref}
\usepackage{arydshln} 
\usepackage[most]{tcolorbox}
\usepackage[T1]{fontenc}
\usepackage{gensymb}
\usepackage{url}

\begin{document}
%
\title{STAA: Spatio-Temporal Alignment Attention for Short-Term Precipitation Forecasting}

\author{Min~Chen,
        Hao~Yang,
        Shaohan~Li,
        and Xiaolin~Qin%
\thanks{This work was supported by the National Key R\&D Program of China (grant no. 2021YFC3000902),  Smart Gridded Forecast Innovation Team Fund of Sichuan Meteor-ological Administration.(Hao Yang and Shaohan Li are co-corresponding authors.)}%
\thanks{Min~Chen and Xiaolin~Qin are with the Chengdu Computer Application Institute, Chinese Academy of Sciences, Chengdu 610041, China; (e-mail:\url{susie1@cuit.edu.cn;qinxl2001@126.com}).}
\thanks{Hao~Yang and Shaohan~Li are with the Chengdu University of Information Technology, Chengdu 610225, China; (\url{e-mail:haoyang@cuit.edu.cn;3220609012@stu.cuit.edu.cn}).}
\thanks{}}%

\markboth{Journal of \LaTeX\ Class Files,~Vol.~13, No.~9, September~2014}%
{Shell \MakeLowercase{\textit{et al.}}: Bare Demo of IEEEtran.cls for Journals}

\maketitle

\begin{abstract}
There is a great need to accurately predict short-term precipitation, which has socioeconomic effects such as agriculture and disaster prevention. Recently, the forecasting models have employed multi-source data as the multi-modality input, thus improving the prediction accuracy. However, the prevailing methods usually suffer from the desynchronization of multi-source variables, the insufficient capability of capturing spatio-temporal dependency, and unsatisfactory performance in predicting extreme precipitation events. To fix these problems, we propose a short-term precipitation forecasting model based on spatio-temporal alignment attention, with SATA as the temporal alignment module and STAU as the spatio-temporal feature extractor to filter high-pass features from precipitation signals and capture multi-term temporal dependencies. Based on satellite and ERA5 data from the southwestern region of China, our model achieves improvements of 12.61\% in terms of RMSE, in comparison with the state-of-the-art methods.
\end{abstract}


%
\IEEEpeerreviewmaketitle

\section{Introduction}
Accurate precipitation forecasting is significant for modern society in many aspects, including transportation planning, natural disaster early warning, new energy and power planning, and so on. Precipitation forecasting tasks can be differentiated into long-term and short-term predictions, based on the differences in the range of prediction time duration.  Forecasts for the next 0-2 hours are  defined as proximity forecasts, and forecasts for the next 2-12 hours are called short-term forecasts according to World Meteorological Organization \cite{schmid2019nowcasting}. 

Deep learning methods have emerged in meteorological fields such as weather forecasting. The improved effectiveness and efficiency enable them to thus become a key research and development focus for major meteorological centers and weather service companies around the world. MetNet \cite{sonderby2020metnet} extends short-term precipitation forecasting to 8 hours and surpasses HRRR \cite{benjamin2016north} and optical flow methods \cite{ayzel2019optical} on this scale.  NowcastNet \cite{zhang2023skilful} achieved state-of-the-art performance in precipitation nowcasting tasks, providing skillful forecasts at light-to-heavy rain rates. Pangu-Weather model \cite{bi2023accurate} for the first time surpassed ECMWF's Integrated Forecasting System \cite{bougeault2010thorpex}, which has the highest accuracy among all the traditional numerical methods, in medium- and long-term weather forecasting.  Further, GraphCast \cite{lam2022graphcast}, outperforms the Pangu-Weather model in medium-range global weather forecasting. 


\begin{figure*}[ht!]  
    \begin{minipage}{0.75\textwidth}
        \includegraphics[width=1\linewidth , trim = 0 0 0 0,clip]{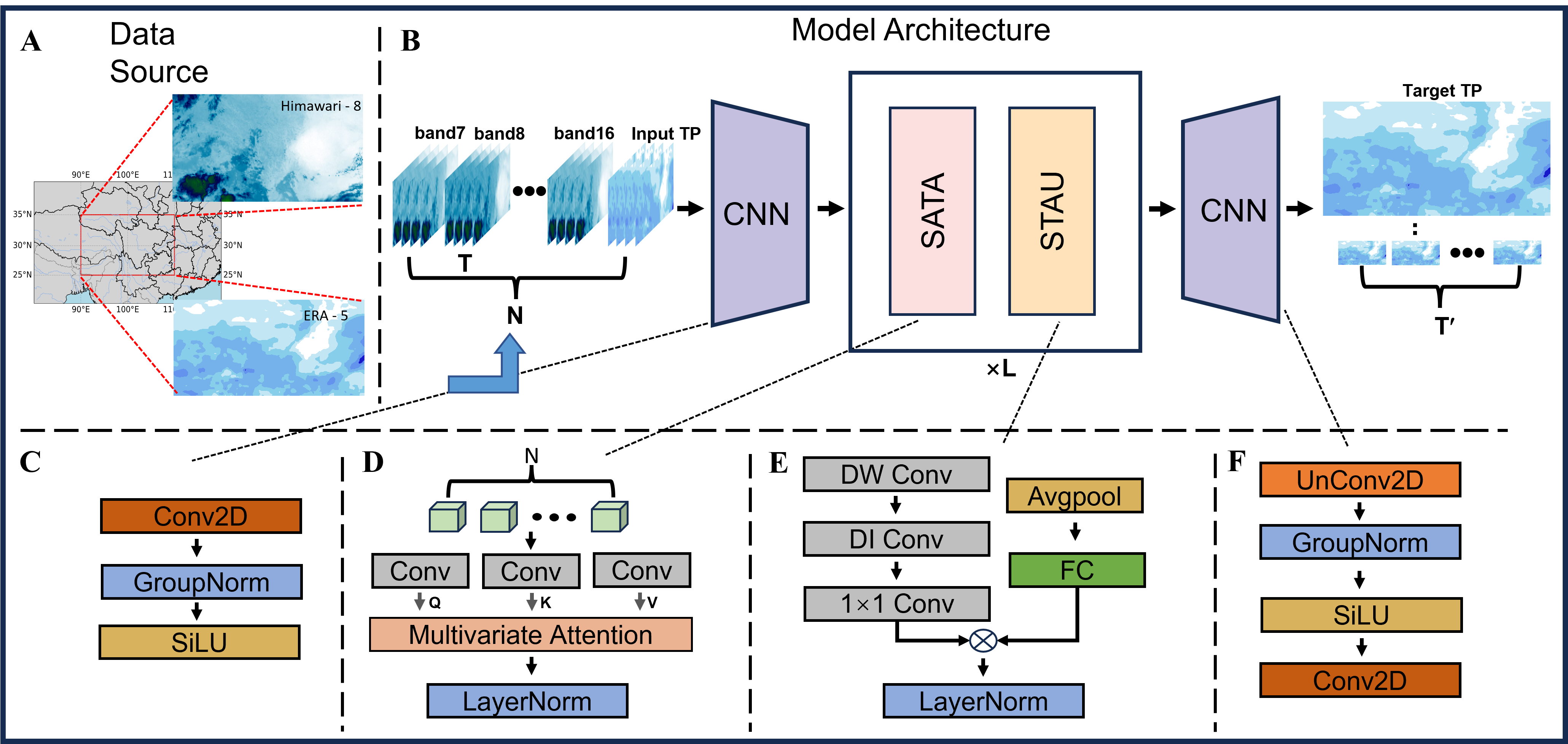}
        \caption{STAA workflows:  First, it encodes spatio-temporal data using Convolution Neural Networks (CNNs) while maintaining the independence of variable channels. Next, a SATA automatically learns the correlations between variables. Subsequently, STAU integrates the spatio-temporal features of various variables. Finally, CNNs are again used to decode the embedding back to the spatio-temporal domain. `$\times L$' means the module are stacked $L$ times.} 
        \label{fig:STAA}
    \end{minipage}  \hspace{1em} \vspace{-1em}  
    \begin{minipage}{0.22\textwidth}
        \centering
        \captionof{table}{A comparative analysis of high-frequency information in hourly TP (total precipitation) data from ERA5 and CIFAR-10\cite{krizhevsky2009learning} image data. `Mean', `Max' and `Min' represent the average, maximum, and minimum values of the strength of the frequency components within the high-frequency domains.}
        \begin{tabular}{clr}
            \hline
            \multicolumn{1}{l}{}
            Data &      & Value  \\ \hline
            \multirow{3}{*}{\rotatebox{90}{\footnotesize{ERA5-TP}}}& Mean & 14.61  \\ 
                                 & Max  & 246.09 \\
                                 & Min  & 0.02   \\
            \midrule
            \multirow{3}{*}{\rotatebox{90}{\footnotesize{CIFAR-10}}}       & Mean & 1.29   \\
                                 & Max  & 45.59  \\
                                 & Min  & 0.00    \\ \hline
            \end{tabular}
        \label{tab:t1}
    \end{minipage}
\end{figure*}
Therefore, deep learning methods are of practical significance for the task of precipitation prediction. Initially, most works employ single radar or satellite images to learn nonlinear functions to realize the prediction \cite{han2023precipitation,gao2022attention}, while precipitation is the result of the interaction of complex environmental factors (e.g., wind, terrain, temperature, barometric pressure, etc.). In recent years, the development of remote sensing technologies has helped to improve the spatial and temporal resolution and accuracy of meteorological observations, which opens up a new space for deep learning methods based on multi-source data fusion. However, challenges still exist in the prevailing methods for multi-source data fusion in short-term precipitation forecasting. Here we aim to focus on three of them: (i) the problems of imprecise temporal alignment of multi-source observation data, (ii) the insufficient representation capability of capturing temporal dependency at different time scales, and (iii) the unsatisfactory prediction performance for sudden changes and extreme events in precipitation.

In this paper, we propose a short-term precipitation multi-step forecasting model named STAA (\underline{s}patio-\underline{t}emporal \underline{a}lignment \underline{a}ttention) based on attention mechanisms in Transformer \cite{vaswani2017attention, liu2023itransformer}. 
Our model adopts a 2D multi-head variable self-attention mechanism to automatically learn the relationships among different sources of variables to extract time-aligned embeddings, we call it SATA (\underline{s}elf-\underline{a}ttention for \underline{t}emporal \underline{a}lignment). Meanwhile, we introduce  STAU (\underline{s}patio-\underline{t}emporal \underline{a}ttention \underline{u}nit) to integrate the spatial features of each variable and capture temporal dependency, which helps to improve the model's representation capability. Large-kernel convolution as a feature of a high-pass filter is used to extract high-frequency features to improve the model's ability to fit sudden changes and extreme events in precipitation.

To evaluate our methods' practical values, we collect ERA-5 \cite{essd-13-4349-2021} and Himawari satellite remote sensing data as the fused multi-source inputs, to obtain rich contextual information for the neural network. Experimentally, STAA shows the best performance in metrics of RMSE, MAE, and PCC. In specific, the RMSE is reduced by 42.20\%, 13.65\%, and 12.61\%, compared to the ConvLSTM, PhyDNet and SimVP models respectively.

\section{Data Source and Study Area}
 In this study, we focus on the southwestern region of China, primarily including provinces of Sichuan, Chongqing, Yunnan, and Guizhou,  as shown in Figure.~\ref{fig:STAA}(A). This region is geographically unique and features a complex terrain structure\cite{DQXK201606009}, and precipitation in this region significantly impacts agricultural production, water resources, and hydroelectric power of the whole country.  However, meteorological stations are often unevenly distributed, especially in remote and high-altitude areas which we focus on. Therefore, we employ ERA5 reanalysis data, with globally consistent high spatial resolution, ensuring continuous and uniform coverage across the southwestern region \cite{xu2024incorporating}. 
 The other modality is satellite infrared signal, which is chosen as Himawari-8 \cite{WXYG201501025}, equipped with state-of-the-art imaging instruments, delivers high temporal resolution data every 10 minutes, and excels in cloud image quality, channel range, spectral bands, and clarity. This satellite is crucial for detecting extreme natural disasters such as heavy rain, dense fog, and typhoons. 

\begin{figure*}[ht!]  
        \begin{minipage}{0.65\textwidth}
        \centering
        \captionof{table}{Results of performance comparison and ablation study. `${\uparrow}$' means the higher the better, and `${\downarrow}$' means the inverse. The values in \textbf{bold} are the top-1 results. The \underline{underlined} values are sub-optimal results. `IMP(\%)' is the percent of improvements of STAA over the sub-optimal ones.  } \vspace{-0.5em} 
        \resizebox{1.0\columnwidth}{!}{
        \begin{tabular}{lrrrrrr}
        \toprule
        Method       & RMSE $({\downarrow})$    & MAE $({\downarrow})$     & PCC $({\uparrow})$     & CSI $({\uparrow})$     & POD $({\uparrow})$    & FAR $({\downarrow})$     \\ 
        \midrule
        ConvLSTM     & 4.83331 & 2.42096 & 0.69594 & 0.61232 & 0.75232 & 0.23308 \\
        PhyDNet      & 3.23541 & \underline{1.59263} & 0.78667 & 0.71028 & \underline{0.85033} & 0.18825 \\
        SimVP        & \underline{3.19666} & 1.60817 & \underline{0.78940} & \underline{0.72040} & 0.84557 & \underline{0.17045} \\
        STAA          & \textbf{2.79365} & \textbf{1.36108} & \textbf{0.80366} & \textbf{0.74122} & \textbf{0.85456} & \textbf{0.15178} \\ 
        \midrule
        IMP(\%) & 12.61\% & 14.54\% & 1.81\% & 2.89\% & 0.50\% & 10.95\% \\
        \midrule
        w/o-SATA & 3.24460 & 1.56339 & 0.80336 & 0.73134 & 0.81817 & 0.12673 \\
        w/o-STAU    & 2.99620 & 1.46208 & 0.77132 & 0.71924 & 0.82822 & 0.15465 \\ \bottomrule
        \label{tab:e1}
        
        \end{tabular}}
        \label{tab:my_label2}
    \end{minipage}\hspace{1em} \vspace{-1em} 
    \begin{minipage}{0.32\textwidth}
        \vspace{-0.5em}
        \includegraphics[width=1\linewidth , trim = 0 00 0 0,clip]{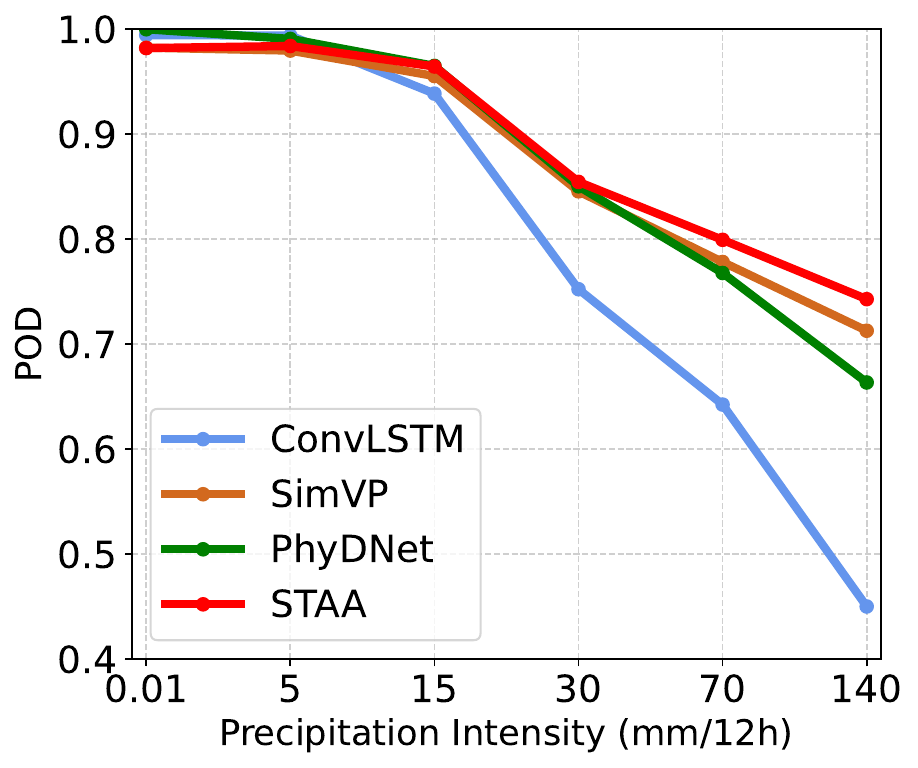} \vspace{-1.4em}
        \caption{Performance of included models on POD metric at different thresholds.} 
        \label{fig:pod}
    \end{minipage} \vspace{-1em}   
\end{figure*}
\section{Methodology}
STAA is an advanced multi-step short-term precipitation prediction model that integrates satellite infrared data and precipitation records. In this process, the task is treated as a spatio-temporal prediction problem. Given that the dataset we utilized, ERA5, has an hourly time resolution, the STAA model accordingly sets each prediction step to one hour. Figure.~\ref{fig:STAA} demonstrates the workflow of the STAA model. 
The model comprises two key components: SATA and STAU. The former used for temporal alignment is achieved by a 2D multi-head self-attention mechanism to automatically learn the correlations between variables so that the extracted embeddings represent aligned features. The latter are spatio-temporal attention units, to integrate the spatial features of variables and capture the multi-term dependencies of the time series. 
\subsection{Self-Attention for Temporal Alignment}
Due to the inherent temporal delays and desynchronization in meteorological observations of different variables, imprecise alignment is usually unavoidable in the time points of different modalities. 
Previous multi-modal methods usually treat the different variables as temporally synchronized, which poses hazards in prediction performance since extraneous noise is likely to be introduced to the feature embeddings due to temporal misalignment. 
 To address the problem, inspired by iTransformer \cite{liu2023itransformer}, we introduce a multi-head self-attention mechanism to learn the temporal dependencies among different variables so that the extracted embeddings are automatically aligned. For a spatio-temporal signal  of the length $T$ and composed of $N$ variables with the spatial resolution of $(H, W)$, which is denoted by $\mathbf{X} \in \mathbb{R}^{N\times T \times H \times W}$, the attention matrix over variable axis reads 
 \begin{equation}
     \mathbf{A} = \mathrm{softmax}(\frac{\mathbf{QK}^\intercal}{\sqrt{D}}),
 \end{equation}
in which $\mathbf{Q}, \mathbf{K} \in\mathbb{R}^{N \times D \times H \times W}$ are query and key, which are extracted by two different 2D-CNNs, and $\sqrt{D}$ is a scaling term. 
After the $\mathrm{softmax}(\cdot)$ operation, $\mathbf{A} \in \mathbb{R}^{N \times N}$ demonstrates the temporal relativity of different variables, and it can be employed to operate time-alignment by $\mathbf{X}_a = \mathbf{A} \mathbf{V}$, where $\mathbf{V} \in \mathbb{R}^{N \times T \times H \times W}$ is the value term, extracted by another 2D-CNN, $\mathbf{X}_a \in \mathbb{R}^{N \times T \times H \times W}$ is the aligned embeddings.


\subsection{Spatio-Temporal Attention Unit}
In the observations, we conclude that the spatial signals in precipitation are of sharper changes, which can be demonstrated by a case study in Table.~\ref{tab:t1}
Based on it, we argue that different from image data, high-frequency components dominate the precipitation data, indicating higher spatial frequencies and more dramatic spatial variations. Hence, the prediction task requires the model to serve as a high-pass filter to extract the spatial features. By this means, we utilize CNNs (Convolutional Neural Networks) as the base feature extractor, which are proven to be effective in extracting high-frequency information \cite{park2022vision}. Besides, we further enhance the models' ability to extract features with large-kernel convolutions which increases the receptive field, since it helps each spatial point to aggregate information from farther points allowing the model to capture the long-range sudden changes.
For computational efficiency, we, inspired by the latest advancements in Vision Transformers (ViTs) and large-kernel convolutions, instead decompose the large-kernel convolution into small-kernel depth-wise convolutions ($\operatorname{Conv}_{\mathrm{DW}}$), depth-wise convolutions with dilation ($\operatorname{Conv}_\mathrm{DI}$), and 1x1 convolutions ($\mathrm{Conv}_{1 \times 1}$). In this way, the spatial attention branch reads:
\begin{equation}
\begin{aligned}
\mathbf{A}_{\mathrm{Sp}}& =\mathrm{Conv}_{1 \times 1}( \operatorname{Conv}_\mathrm{DI}( \operatorname{Conv}_{\mathrm{DW}}(\mathbf{X}_a))),
\end{aligned}
\end{equation}
where $\mathbf{X}_a$ represents the output of the SATA module, $\mathbf{A}_{\mathrm{Sp}} \in \mathbb{R}^{N \times T \times H \times W}$ denotes the spatial attention weight matrix.

To enhance the model's medium-to-long-term forecasting ability in the temporal domain, we focused on improving the model's capability to capture long-term dependencies in time series, by introducing a dynamic attention mechanism. We employ a squeeze-and-excitation strategy to learn attention weights over the temporal channels, thereby effectively capturing long-range dependencies in the time series, which leads to the temporal attention branch in STAU, as
\begin{equation}
\begin{aligned}
\mathbf{A}_{\mathrm{Tp}} & =\mathrm{FC}(\operatorname{AvgPool}(\mathbf{X}_a))),
\end{aligned}
\end{equation}
where $\mathbf{A}_{\mathrm{Tp}} \in \mathbb{R}^{N\times T \times 1 \times 1}$ is the temporal attention, $\mathrm{FC}(\cdot)$ and $\operatorname{AvgPool}(\cdot)$ are fully-connected layers and average pooling.

Finally, STAU is composed of two branches, SA (Spatial Attention) and TA (Temporal Attention), as shown in Appendix.~\ref{app:method}, and the final aggregation of spatial and temporal information reads
\begin{equation}
\begin{aligned}
\mathbf{X}_o & =(\mathbf{A}_{\mathrm{Sp}} \otimes \mathbf{A}_{\mathrm{Tp}}) \odot \mathbf{X}_a,
\end{aligned}
\end{equation}
where $\mathbf{X}_o \in \mathbb{R}^{N \times D \times H \times W}$ represents the output of the STAU, $\otimes$ is the Kronecker product and  $\odot$ is the Hadamard product.

\subsection{Training}
Finally, another CNN (Convolutional Neural Network) is used to decode the embedding back to the spatio-temporal domain of precipitation, as shown in Fig.~\ref{fig:STAA}. 
It is composed of deconvolution layers, group normalization, SiLU activation functions, and Conv2D layers stacked together. Mean Square Error (MSE) loss is used to train the model, leading to the training objective as
\begin{equation}
\begin{aligned}
\mathcal{L} = || \mathrm{Dec}(\mathbf{X}_o) - \mathbf{\hat{X}}||_2^2,
\end{aligned}
\end{equation}
where $\mathbf{\hat{X}} \in \mathbb{R}^{N \times T' \times H \times W}$ is the ground truth observations, $\mathrm{Dec}(\cdot)$ is the CNN-based decoder.
\section{Experiment}
\begin{figure}[h]
     \centering
    \includegraphics[width=0.9\columnwidth, trim = 10 0 0 0]{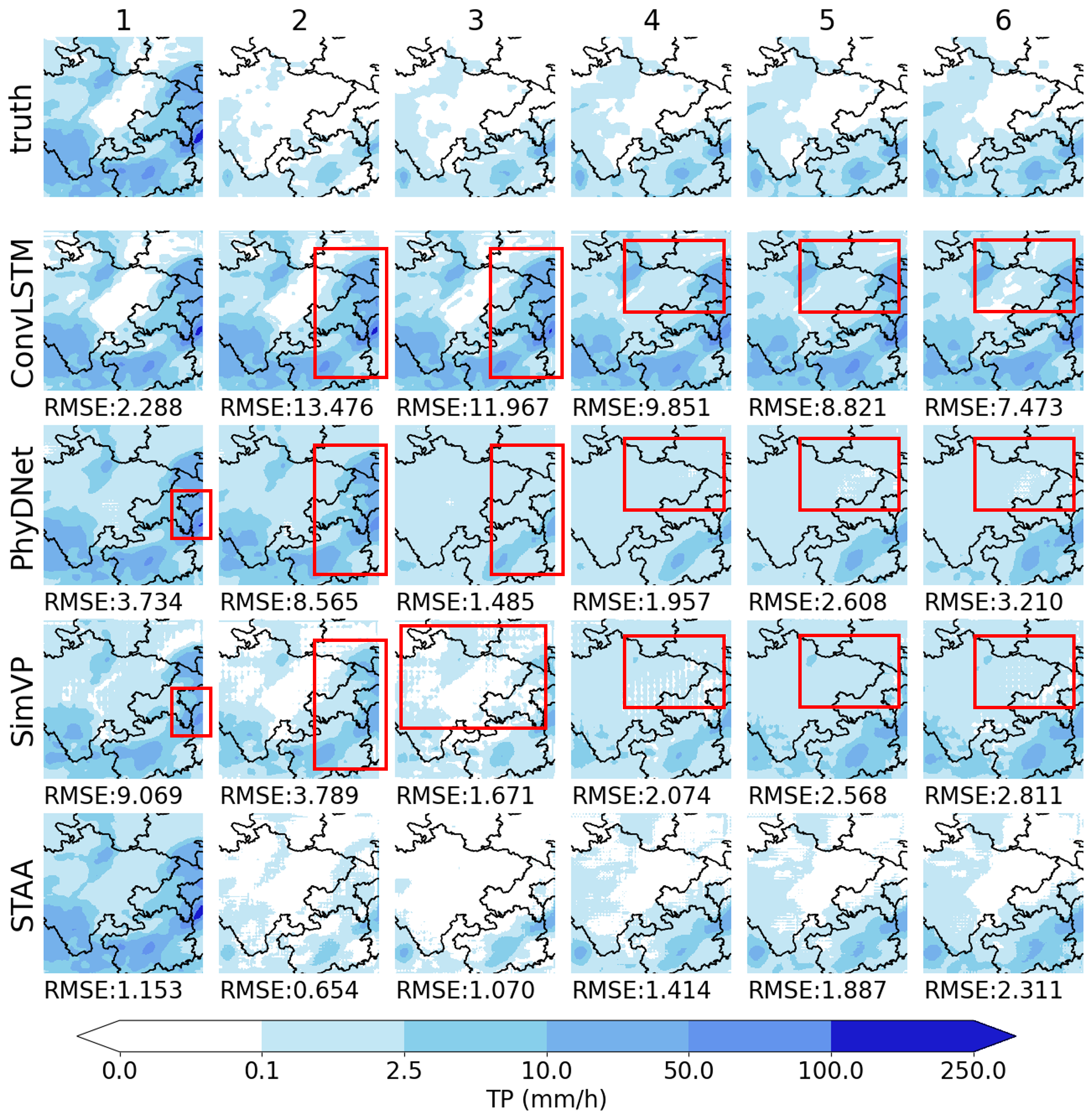}
    \caption{Comparison of different models in hourly precipitation prediction as a case study. The red frame demonstrates the deficits in the prediction of different methods.}
    \label{fig:e1} \vspace{-1em}
\end{figure}
\subsection{Dataset}
In this study, we use Himawari satellite data and ERA5-LAND precipitation data from 2017 to 2021 in flood seasons. Specifically, the flood season refers to the period from May to September each year, and the validation set consists of data from June to July 2021, the test set consists of August and September data in the same year, and the remaining data constitutes the training set. The training data includes 8360 consecutive hours of observations, 1459 and 1460 hours of observations make up the validation and test sets, respectively. The precipitation forecasting task can be considered as a spatio-temporal prediction task, in which we follow the setting in CLCRN \cite{lin2021conditional}, with the past 12-hour spatial observations of precipitation and satellite as input, to predict the future 12-hour precipitation (Hourly forecasting).
The global observations are cropped, with the spatial region ranging from 25\degree-35\degree N and 90\degree-110\degree E, with a grid count of 101$\times$201. We adjust the spatial and temporal resolution of the Himawari satellite data to coincide with the temporal and spatial resolution of ERA5, which is 0.1\degree and 1h.  Therefore, the spatial grid size in this study is set to 101 $\times$ 201, and the number of input and output time steps are both 12. Finally, the 7-16 bands from the Himawari satellite and precipitation are concatenated in the channel dimensions, leading the input channels to equal 11.
\subsection{Implementation Details}
The concatenated spatio-temporal data as input into the encoder, is firstly transformed into high-dimensional embeddings; and then decoded as the predicted precipitation. In the training stage, Adam serves as the optimizer. The learning rate decay scheduler is adopted, in which the initial learning rate is set to 0.0001 and the decaying coefficient is 0.5. The batch size is set to 6, the max training epoch number is set to 200, and the early stopping strategy is employed.
\subsection{Evaluation Metrics}
To evaluate the performance of STAA quantitatively, we adopt six widely-used metrics: \textbf{RMSE} as root mean square error, \textbf{MAE} as mean absolute error, \textbf{PCC} as Pearson correlation coefficient, \textbf{CSI} as critical success index, \textbf{POD} as probability of detection and \textbf{FAR} as false alarm ratio. 
In calculating CSI, POD, and FAR, a threshold is used to convert each grid point of the prediction results into binary values. The detailed calculation methods can be found in the Appendix.~\ref{app:A}.

\subsection{Results Comparison}
To validate the effectiveness of the proposed method, we choose three state-of-the-art methods on spatio-temporal prediction tasks for comparison, including ConvLSTM \cite{shi2015convolutional}, PhyDNet \cite{guen2020disentangling} and SimVP \cite{gao2022simvp}. 
Table.~\ref{tab:e1} gives the results of different methods according to average metrics over the 12 hours. It can be concluded that (1) STAA shows the best performance in metrics of RMSE, MAE and PCC. In specific,  
The RMSE is reduced by 42.20\%, 13.65\% and 12.61\%, compared to the ConvLSTM, PhyDNet and SimVP models respectively.
(2)  CSI, POD and FAR as binary classification metrics, measure the correctness of the model in predicting heavy rainfall. We found that the STAA demonstrated the best performance on the CSI, POD and FAR  metrics, indicating that it is capable of predicting extreme rainfall events, and also, ability to fit the sudden changes in signals. Overall, our model succeeds in achieving significant improvements in forecasting accuracy in comparison with the state-of-the-art methods.

Here, we further give a case study on of the first 6 hours in 2021-08-24, 25\degree-35\degree N and 100\degree-110\degree E in Fig.~\ref{fig:e1}. The red frame demonstrates the deficits in the prediction of different methods. In the initial phases of the prediction, particularly in the first hour, it is noted that an extreme precipitation value (TP > 100mm/h) emerged in the east of the selected region. The red-boxed areas show that ConvLSTM and STAA can capture the extreme value in spatial domains, whereas the PhyDNet and SimVP hardly forecast this extreme value. STAA not only precisely captures the extreme events in the area, but also depicts the distribution of the precipitation in the overall region, from its smaller predicted RMSE in the first hour than ConvLSTM.
In the next two hours, the precipitation patterns in the region undergo a significant change, with precipitation diminishing rapidly or even being absent in certain areas. The sudden diminishment is only detected by the STAA with high predictive accuracy, whose forecasting results closely align with the actual conditions in most areas. SimVP also detects the change but tends to overestimate the areas, particularly in the red-boxed areas. Conversely, the ConvLSTM and PhyDNet fail to foresee this event. Notably,  STAA markedly surpasses all the baselines in the subsequent hours, since it succeeds in capturing the regional spreading of precipitation from the southwest to the northeast. 

\begin{figure*}[h]
    \centering
    \includegraphics[width=0.9\linewidth]{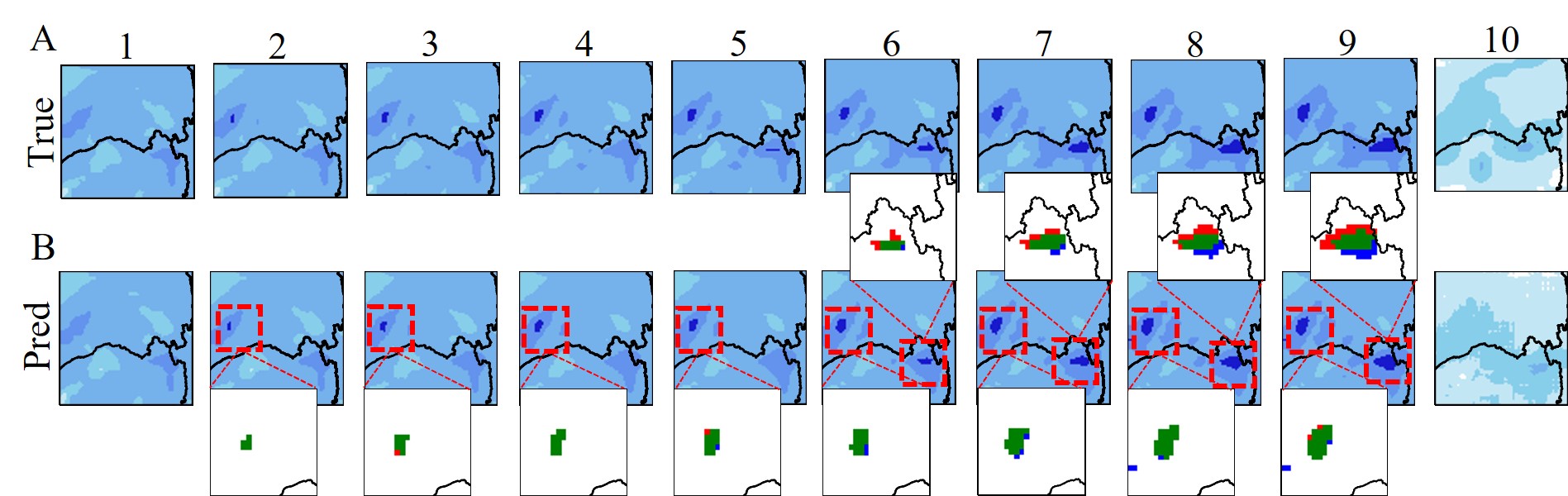}
    \caption{An analysis of the case from 03:00 to 13:00 on August 25, 2021, within the region of 26\degree-31\degree N and 94\degree-99\degree E. A: Actual observations. B: STAA model prediction results. Numbers 1-10 represent specific time periods in this case study. For heavy rain prediction (accumulated rainfall reaching or exceeding 100 mm/hour), green indicates correct predictions, blue indicates false alarms, and red indicates missed predictions. } 
    \label{fig:eg}\vspace{-1em}
\end{figure*} 
\subsection{Further Analysis}        
To elucidate the patterns that STAA can forecast with high accuracy, we assess the models' performance using the POD metrics across different precipitation intensities as shown in Fig.~\ref{fig:pod}. At a precipitation intensity of 0.01, the metrics of various models show no significant differences, indicating similar rainfall prediction capabilities. The performance of the ConvLSTM model drops sharply when the precipitation intensity reaches 30 mm/12 hours. And, as the precipitation intensity increases, the gap between the metric curve of the STAA model and other models widens, demonstrating that the STAA model has a clear advantage in predicting high precipitation intensities.

\subsection{Case Study} 
In Figure.~\ref{fig:eg}, we present a case study that encompasses the entire heavy rainstorm event, with accumulated precipitation of 100 mm/h or more, which occurred from 03:00 to 13:00 on August 25, 2021, within the region of 26\degree-31\degree N and 94\degree-99\degree E. The extreme rainstorm (accumulated precipitation of 100mm/h or more) began in the second hour and ended in the tenth hour. The STAA model accurately predicted this time interval. 
Additionally, the color distribution in Figure .~\ref{fig:eg}B further verifies the effectiveness of the STAA model in predicting extreme rainfall events. The prevalence of green indicates a broad range of areas where predictions align with actual conditions, while the lower presence of red and blue signifies fewer missed and false alarms. 
The accurate prediction of heavy rainfall in terms of time and space highlights the efficiency and accuracy of the STAA model in detecting and predicting extreme weather events. This high-precision temporal and spatial prediction is crucial in practical applications as it allows relevant departments to prepare in advance, thereby minimizing potential damage caused by heavy rainfall. 

\subsection{Ablation Analysis}
We conducted ablation experiments on two major modules in STAA. Table.~\ref{tab:e1} reveals the performance gain brought about by the SATA and STAU modules. In detail, eliminating the SATA module increases the Root Mean Square Error (RMSE) by 16.14\%. Similarly, removing the STAU module raises the RMSE, leading to a 7.25\% increase in RMSE. These results underscore the importance of the SATA and STAU modules. A more detailed case study is given in appendix Appendix.~\ref{app:B}. 
\section{Conclusion}
We proposed STAA which achieves improved performance in the multi-step prediction of short-term precipitation. The approach is based on the attention mechanism, utilizing a two-dimensional multi-head variable self-attention mechanism (SATA) and a spatial-temporal attention unit (STAU).
In practice, we employ data from ground weather stations and Himawari Satellite as multi-source observation data. Future work will focus on optimizing the model architecture, injecting physics information into the model, and exploring the possibility of adapting the model to more meteorological tasks.


%





%
\bibliographystyle{IEEEtran}
\bibliography{bibtex/bib/myReferences}



%








\newpage
\appendices
\section{}
\label{app:data_statistics}
\begin{table}[h]
\captionof{table}{Mean of Himawari satellite data bands 7-16 and ERA5 precipitation data from 2017 to 2021} \vspace{-0.5em} 
\begin{tabular}{lrrrrr}
\toprule
data            & 2017 & 2018 & 2019 & 2020 & 2021 \\ 
\midrule
h08\_band7  & 275.6185                 & 275.4190                  & 236.0102                 & 222.0135                 & 182.0661                 \\
h08\_band8  & 232.1687                 & 231.8001                 & 192.5284                 & 178.8094                 & 138.9822                 \\
h08\_band9  & 239.4687                 & 239.1535                 & 200.0398                 & 186.2370                  & 146.5240                  \\
h08\_band10 & 245.5260                  & 245.2971                 & 206.0584                 & 192.1624                 & 152.3500                   \\
h08\_band11 & 261.5767                 & 261.2239                 & 222.0605                 & 207.5874                 & 167.9538                 \\
h08\_band12 & 246.6595                 & 246.6097                 & 206.2194                 & 193.3991                 & 152.4282                 \\
h08\_band13 & 263.3518                 & 262.9748                 & 223.8726                 & 208.9850                  & 169.4631                 \\
h08\_band14 & 262.2831                 & 261.8884                 & 222.7816                 & 207.7322                 & 168.1824                 \\
h08\_band15 & 260.2146                 & 259.8186                 & 221.3440                  & 206.4145                 & 168.0741                 \\
h08\_band16 & 252.1270                  & 251.8259                 & 213.1297                 & 198.8317                 & 160.2646                 \\
ERA5\_TP      & 4.4382                 & 3.8958                 & 3.4893                  & 4.4763                 & 3.9432                  \\ \bottomrule
\end{tabular}
\label{tab:t3}
\end{table}
\begin{figure}[h]
     \centering
    \includegraphics[width=1\columnwidth, trim = 20 0 0 0]{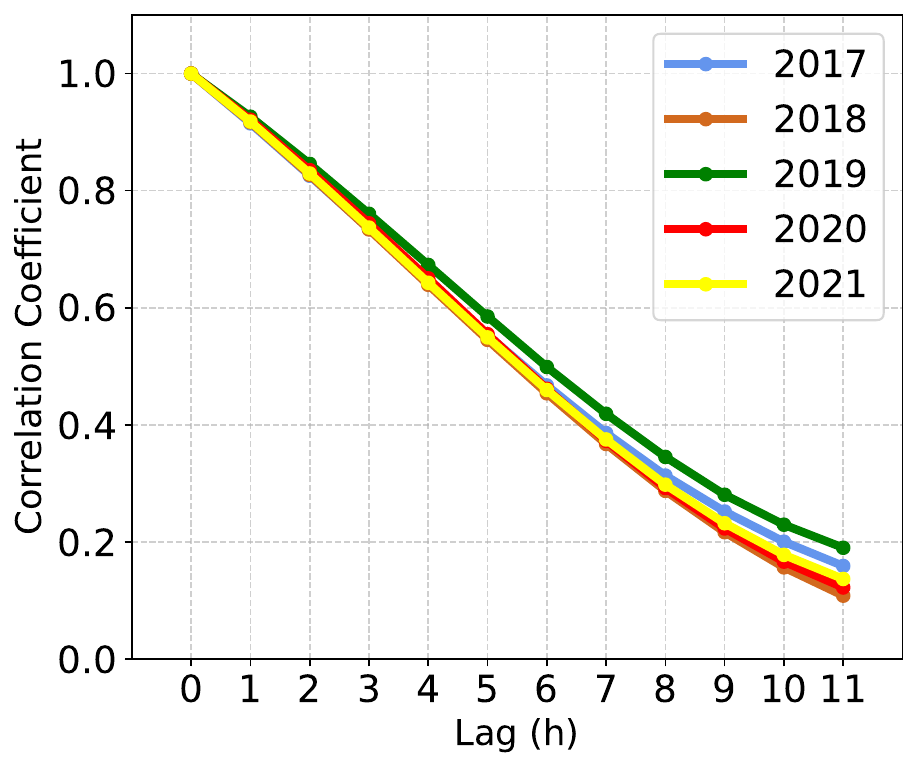}
    \caption{ERA5-TP data lag correlation analysis for the flood seasons of 2017-2021.}
    \label{fig:TP-ANA} \vspace{-1em}
\end{figure}
To gain a better understanding of the dynamic characteristics of the data and to inform the choice of research methods, this paper provides an overview of the dataset used and analyzes its time series characteristics. Table.~\ref{tab:t3} displays the basic information of the dataset, while Figure.~\ref{fig:TP-ANA} illustrates the lag correlation analysis of ERAT-TP data from May to September for each year from 2017 to 2021. The figure indicates that as the lag time increases, the correlation coefficient gradually decreases. This means that the correlation between the data weakens over time. This trend is consistent across all years, although the rate of decline and the correlation coefficient values vary slightly from year to year. By a lag time of 11 hours, the correlation coefficient has dropped to around 0.1, indicating a very low correlation. This is the reason we chose to use historical data from the past 12 hours.

\section{}
\label{app:method}
\begin{figure*}[t]
    \centering
    \subfigure[STAU workflow]{
    \label{Fig.sub.1}
    \includegraphics[height = 4.0cm]{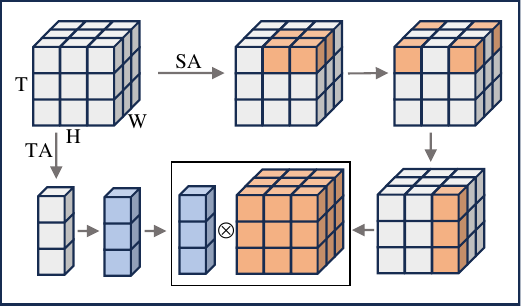}}\hspace{1em}\subfigure[ERA5 precipitation frequency analysis]{
    \label{Fig.sub.2}
    \includegraphics[height = 4.0cm]{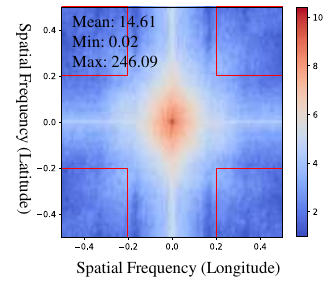}}\hspace{1em}\subfigure[CIFAR-10 frequency analysis]{
    \label{Fig.sub.3}
    \hspace{-2em}\includegraphics[height = 4.0cm]{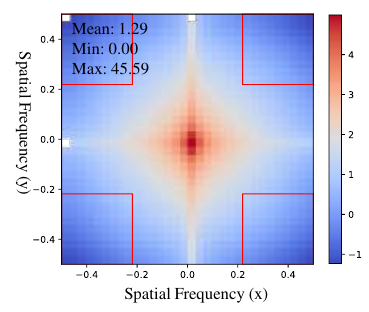}}
    \caption{(a): STAU module architecture: `SA' means the spatial attention module and `TA' means the temporal attention module.
    (b) and (c) display the mean-centered spectral maps of the two-dimensional Fourier transform for precipitation data ranging from 00:00 on August 1, 2021, to 23:00 on August 31 and image data of CIFAR-10 \cite{krizhevsky2009learning}, respectively. The red regions are defined as high-frequency domains in the spatial domain. `Mean', `Max' and `Min' represent the average, maximum, and minimum values of the strength of the frequency components within the high-frequency domains.} \vspace{-1em}
    \label{p2}
\end{figure*}
The detailed architecture of the STAU module and the detailed spectral analysis of ERA5-TP data and CIFAR-10 are provided here, as shown in Figure.~\ref{p2}.
We employ Fourier Transform to analyze the frequency components in the data and apply centralization (shifting low-frequency signals to the center of the spectrum, with high-frequency signals distributed towards the edges). This approach places the low-frequency components (representing the main or average trends of the data) at the center, while the high-frequency components (such as details and rapid changes) are displayed around the periphery, effectively highlighting the frequency characteristics and details of the data. The results are illustrated in Figure.~\ref{Fig.sub.2} and \ref{Fig.sub.3}. Additionally, we calculate the mean, maximum, and minimum values within the red-boxed areas in each figure to intuitively compare the proportion of high-frequency information between the precipitation data and image data.
The figure indicates that the high-frequency components are more prevalent in the precipitation data compared to the image data.
\section{}
\label{app:A}
These metrics at single time step $t$ are defined as below:
\begin{equation}
\begin{aligned}
RMS{E_{\rm{t}}} &= \sqrt {\frac{1}{N}{{\sum\limits_{{\rm{i}} = 1}^N {\left( {{{{\rm{\hat y}}}^{\rm{i}}}_{\rm{t}} - {\rm{y}}_{\rm{t}}^{\rm{i}}} \right)} }^2}} \\
{MAE}_t&=\frac{1}{N} \sum_{i=1}^{N}\left|\hat y_{t}^i-y_{t}^i\right|\\
PCC_t &= \frac{{\sum\limits_{i = 1}^N {\left( {y_t^i - {{\bar y}_t}} \right)}\left( {\hat y_t^i - {{\bar \hat y}_t}} \right)}}{{\sqrt {\sum\limits_{i = 1}^N {{{\left( {y_t^i - {{\bar y}_t}} \right)}^2}}  \cdot \sum\limits_{i = 1}^N {{{\left( {\hat y_t^i - {{\bar \hat y}_t}} \right)}^2}} } }} \\
CS{I_t}&=\frac{TP}{TP + FP + FN}\\
POD_t&=\frac{TP}{TP + FN}\\
FAR_t&=\frac{FP}{TP + FP}
\end{aligned}
\end{equation}
where ${{{\rm{\hat y}}}^{\rm{i}}}_{\rm{t}}$ and ${\rm{y}}_{\rm{t}}^{\rm{i}}$ represent the predicted value and the true value at time t, respectively.
We use a threshold to convert each grid point value in the prediction results into binary values for the calculation of CSI, POD and FAR, which is set as 30mm in practice. In the field of meteorology, when the accumulation of precipitation reaches 30mm in 12 hours, the phenomenon is defined as a Torrential rain. The specific definitions are provided in Table.~\ref{tab:t4}.
TP, FP and FN represent numbers of true positive, false positive, and false negative predictions on rainstorms, respectively. These metrics reflect the performance of models for heavy rainfall prediction.
\begin{table}[]
\centering
\caption{Classification of 12-hour rainfall levels.}
\begin{tabular}{ll}
\hline
\multicolumn{1}{c}{Level}        & \multicolumn{1}{c}{Precipitation (mm/12h)} \\ \hline
Light rain                       & \textless{}5                               \\
Moderate rain                    & 5-15                                       \\
Heavy rain                       & 15-30                                      \\
Torrential rain                  & 30-70                                      \\
Severe torrential rain           & 70-140                                     \\
Extremely Severe Torrential Rain & \textgreater{}140                          \\ \hline
\end{tabular}
\label{tab:t4}
\end{table}

\begin{figure*}[h]
    \centering
    \includegraphics[width=1\textwidth]{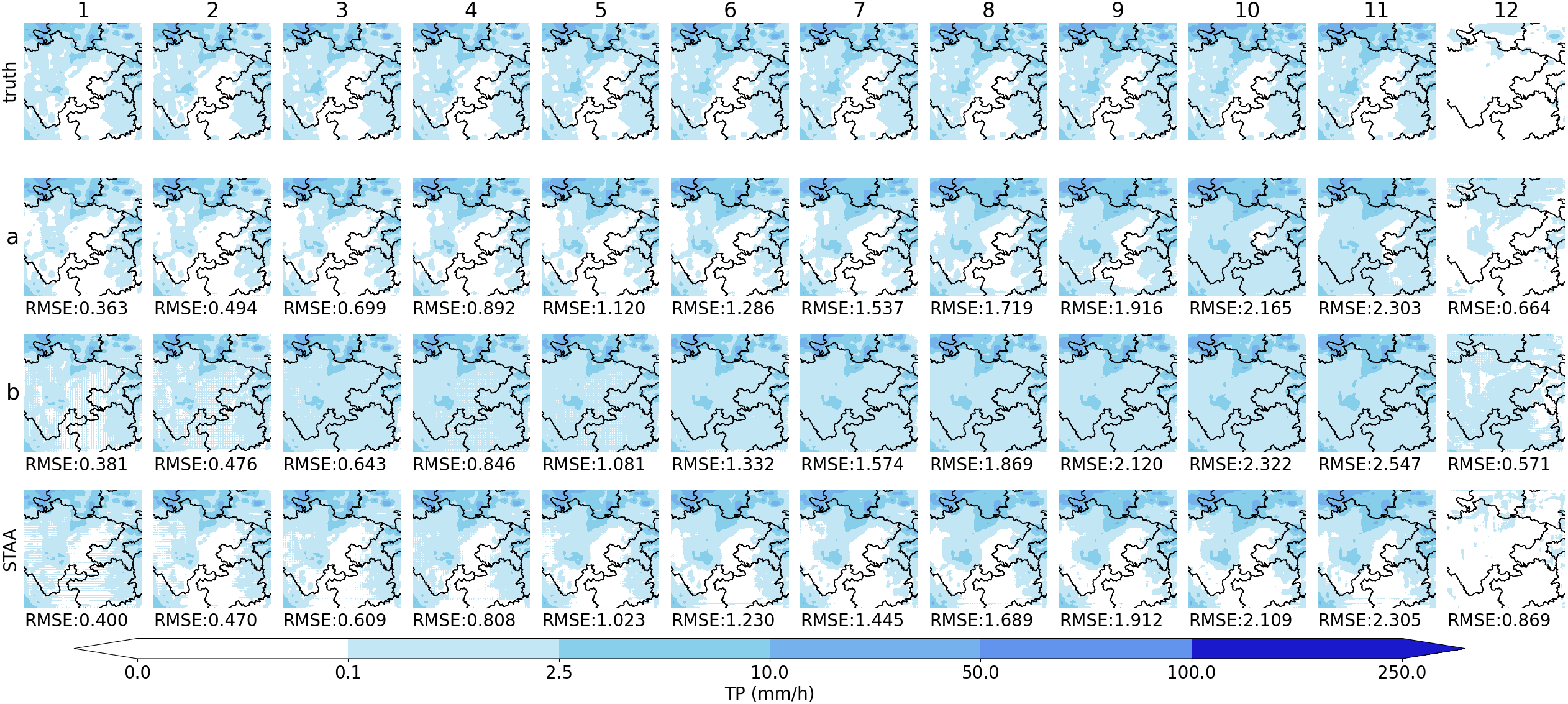}
    \caption{Visualization of module ablation experiments within the 25\degree-35\degree N, 100\degree-110\degree E range.(a represents without SATA;b represents without STAU)}
    \label{fig:e2}
\end{figure*}
\begin{figure*}[h]
    \centering
    \includegraphics[width=1\linewidth]{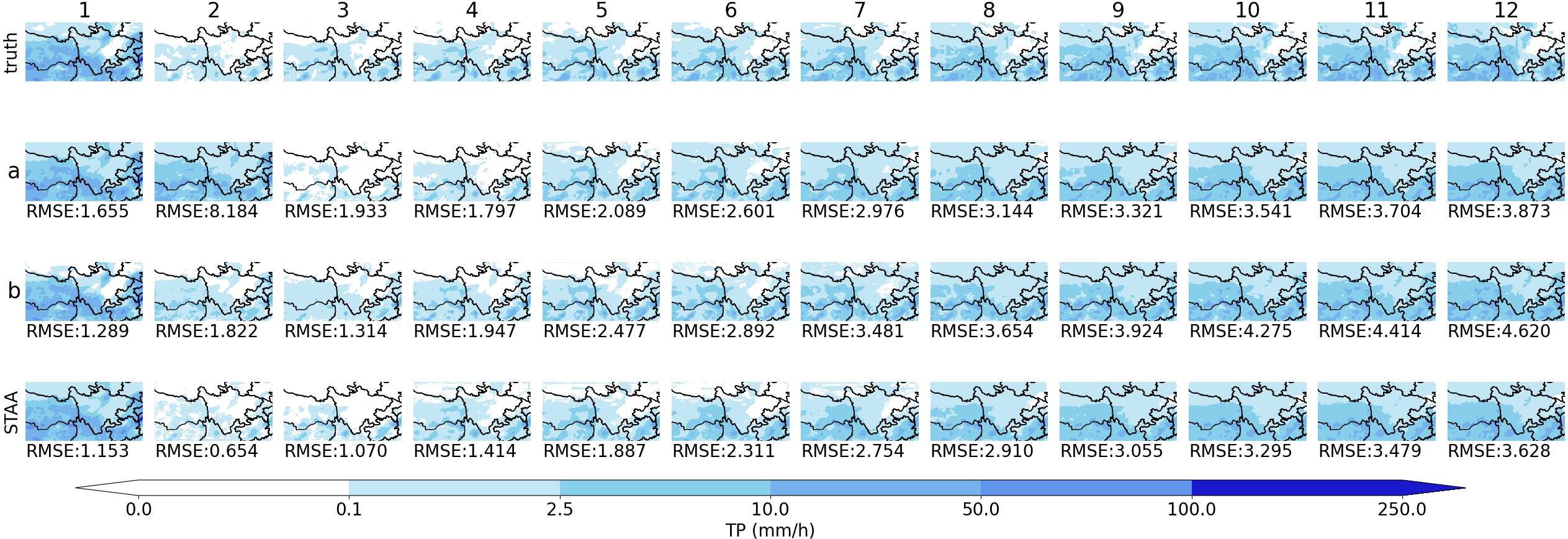}
    \caption{Visualization results of module ablation experiments across the entire study area.(a represents without SATA;b represents without STAU)} 
    \label{fig:e3}
\end{figure*}
\section{}
\label{app:B}
To elucidate the specific impacts of the two modules in depth, we visualized the predictive results of the model after the removal of these modules. Through this intuitive approach, we were able to clearly demonstrate the concrete effects of module ablation on the predictive capability of the model.Figure.~\ref{fig:e2} displays the visual representation of the model's predictive outcomes within the specific geographical region defined by the latitude range of 25\degree-35\degree and the longitude range of 100\degree-110\degree, following the removal of individual modules. From the visualization results in Figure.~\ref{fig:e2}, we can observe that within the initial 1 to 2 hours, regardless of whether the SATA module or the STAU module is removed, the model appears to predict the areas of lower rainfall in the region quite well. However, as the prediction time extends, the model without the STAU module begins to significantly lose its predictive ability for low rainfall areas. In contrast, the model that retains the STAU module continues to maintain predictive accuracy for the region even within a 12-hour forecast period. This observation highlights the importance of the STAU module within the model, particularly in its capacity to capture and process long-term temporal dependencies.

Figure.~\ref{fig:e3} provides a comprehensive perspective, displaying the visualized results of the model's predictive capabilities for the entire study area after the removal of individual modules. From the visualization results in Figure 5, we can distinctly observe a significant reduction in the precipitation across the entire study area during the second hour of prediction. This observation indicates that the model with the SATA module is capable of accurately and promptly capturing such sudden changes in precipitation, successfully predicting this shift as early as the second hour. In contrast, the response of the model to this abrupt change is delayed when the SATA module is removed, with the significant decrease in precipitation only starting to be reflected in the third hour. This finding indeed highlights the pivotal role played by the SATA module within the model, particularly in its capacity to swiftly detect and respond to abrupt changes in climatic conditions. The SATA module, by integrating satellite infrared data and performing multimodal automatic temporal alignment, significantly enhances the model's predictive accuracy for sudden meteorological events. This capability enables the model to better understand and handle complex climate dynamics, especially when predicting rapid changes that may occur in the short term.
\section{}
\label{app:trainvail}
\begin{figure}[h]
     \centering
    \includegraphics[width=0.9\columnwidth, trim = 10 0 0 0]{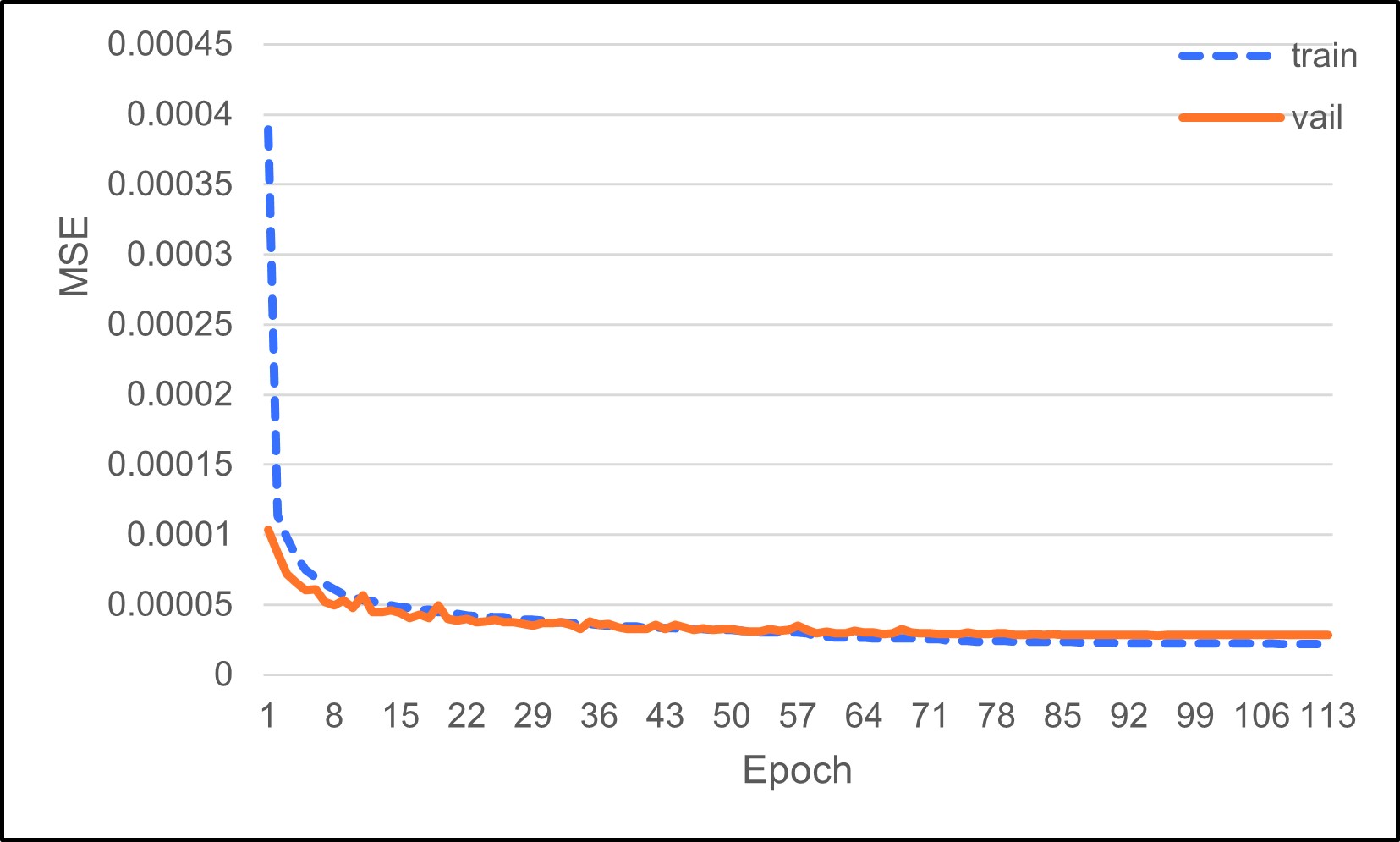}
    \caption{The plot of training set and validation set MSE changes with epochs.}
    \label{fig:trainvail} \vspace{-1em}
\end{figure}
To detail the performance changes of our model during the training process, the following provides the variation of mean squared error (MSE) for the training set and validation set. This data helps us evaluate the adequacy of model training and its generalization ability on different datasets.
Figure.~\ref{fig:trainvail} shows the changes in the training and validation sets over the number of training epochs. It can be observed that the MSE of both the training and validation sets tend to level off towards the end, indicating that our training time is sufficient. Additionally, the MSE for both the training and validation sets is relatively low.

\begin{figure}[h]
     \centering
    \includegraphics[width=0.9\columnwidth, trim = 10 0 0 0]{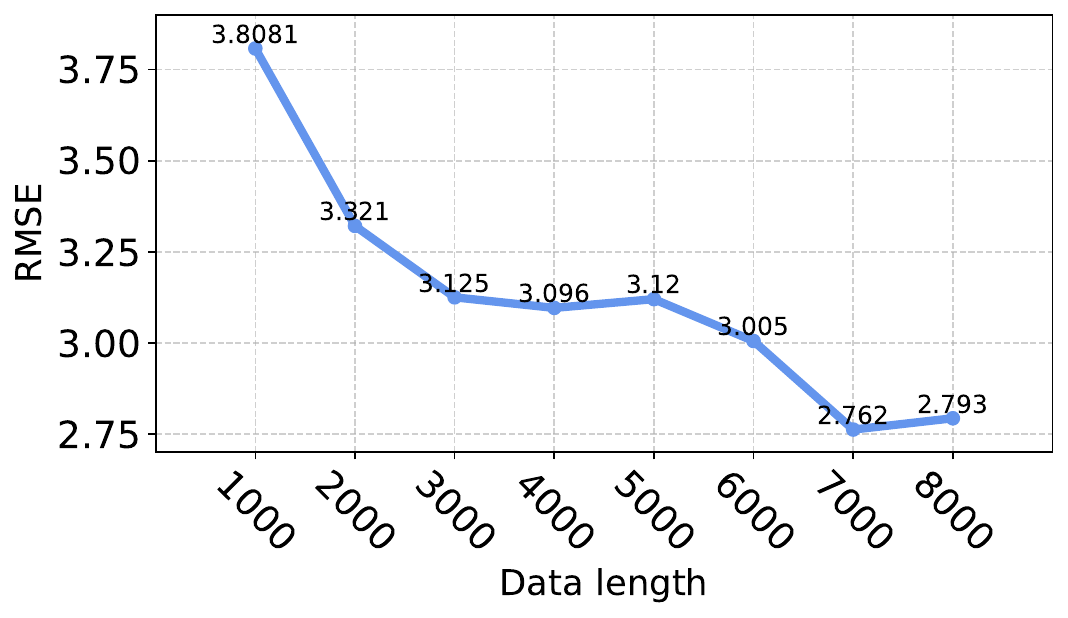}
    \caption{The impact of training data length on prediction accuracy.}
    \label{fig:datalen} \vspace{-1em}
\end{figure}
To find the appropriate training data length, we analyzed the effect of different data lengths on the RMSE of the prediction model. The horizontal axis represents the data length, ranging from 1000 to 8000, while the vertical axis shows the corresponding RMSE values. It can be clearly observed from the graph that as the data length increases, the RMSE value generally decreases, especially after the data length exceeds 5000, where the rate of decrease becomes more gradual.
This trend indicates that adding more data positively impacts improving the prediction accuracy of the model, particularly in the initial stages of data growth. However, when the data length reaches a certain magnitude, the marginal benefit of adding more data for accuracy improvement diminishes. This suggests that the data length we are using is already sufficient, and further increasing the data volume has limited contribution to enhancing model performance.
Through this analysis, we can conclude that the current data volume is reasonable for this prediction model, ensuring its effectiveness and efficiency. While further increasing data length may slightly improve accuracy, considering the cost and benefit, the existing data length is sufficient to meet the model training requirements.
\end{document}